\documentclass[fontsize=9pt,twocolumn,abstract=true]{scrartcl}
\usepackage[detect-weight=true, detect-family=true]{siunitx}
\usepackage{physics}
\usepackage{amsmath}
\usepackage{layouts}
\usepackage{authblk}
\usepackage{graphicx}
\usepackage{microtype}

\title{High-energy ultraviolet dispersive-wave emission in compact hollow capillary systems}

\author[1, *]{Christian Brahms}
\author[1]{Teodora Grigorova}
\author[1]{Federico Belli}
\author[1]{John C. Travers}

\affil[1]{School of Engineering and Physical Sciences, Heriot-Watt University, Edinburgh, EH14 4AS, UK}

\affil[*]{Corresponding author: c.brahms@hw.ac.uk}

\begin{document}

\maketitle
\textbf{
    We demonstrate high-energy resonant dispersive-wave emission in the deep ultraviolet (218 to \SI{375}{\nm}) from optical solitons in short (15 to \SI{34}{\cm}) hollow capillary fibres. This down-scaling in length compared to previous results in capillaries is achieved by using small core diameters (100 and \SI{150}{\micro\meter}) and pumping with \SI{6.3}{\fs} pulses at \SI{800}{\nm}. We generate pulses with energies of 4 to \SI{6}{\micro\joule} across the deep ultraviolet in a \SI{100}{\micro\meter} capillary and up to \SI{11}{\micro\joule} in a \SI{150}{\micro\meter} capillary. From comparisons to simulations we estimate the ultraviolet pulse to be 2 to \SI{2.5}{\fs} in duration. We also numerically study the influence of pump duration on the bandwidth of the dispersive wave.
}
Optical soliton dynamics, obtained by balancing linear and nonlinear contributions to the phase of a propagating light pulse, are a key phenomenon in nonlinear fibre optics. Resonant dispersive wave (RDW) emission in gas-filled hollow fibres is a particularly promising application of this effect, enabling the generation of tuneable ultrashort pulses and supercontinua at shorter wavelengths than possible in any solid-core waveguide \cite{joly_bright_2011, kottig_generation_2017, belli_vacuum-ultraviolet_2015, ermolov_supercontinuum_2015, travers_high-energy_2019}, from the vacuum ultraviolet to the visible spectral region. These dynamics were pioneered in hollow-core photonic-crystal fibres (HC-PCF). Recently, we demonstrated that they can be scaled in energy by up to several orders of magnitude in simple hollow capillary fibres (HCF) \cite{travers_high-energy_2019}. Pumping gas-filled large-core HCF with \SI{10}{\fs} pulses enabled soliton self-compression to \SI{1}{\fs} in the near infrared---an optical attosecond pulse---and the generation of few-femtosecond vacuum and deep ultraviolet (VUV/DUV) pulses at unprecedented peak power, comparable to free-electron lasers. We also showed that up-scaling of those dynamics to the terrawatt scale is feasible by further increasing the HCF core size. Here, we down-scale the core size instead to achieve a more compact and practical setup. Whereas our first demonstration made use of \SI{3}{\m} HCF, here we show that this can be reduced to just 15 to \SI{34}{\cm} when using small-core HCF and even shorter pump pulses---\SI{6.3}{\femto\second}, as readily generated in widely used conventional HCF pulse compression systems. The required pulse energy is also reduced, which is an additional advantage if multiple frequency conversion schemes are to be driven simultaneously, as for instance in multi-colour time-resolved spectroscopy experiments.

\begin{figure*}
    \includegraphics{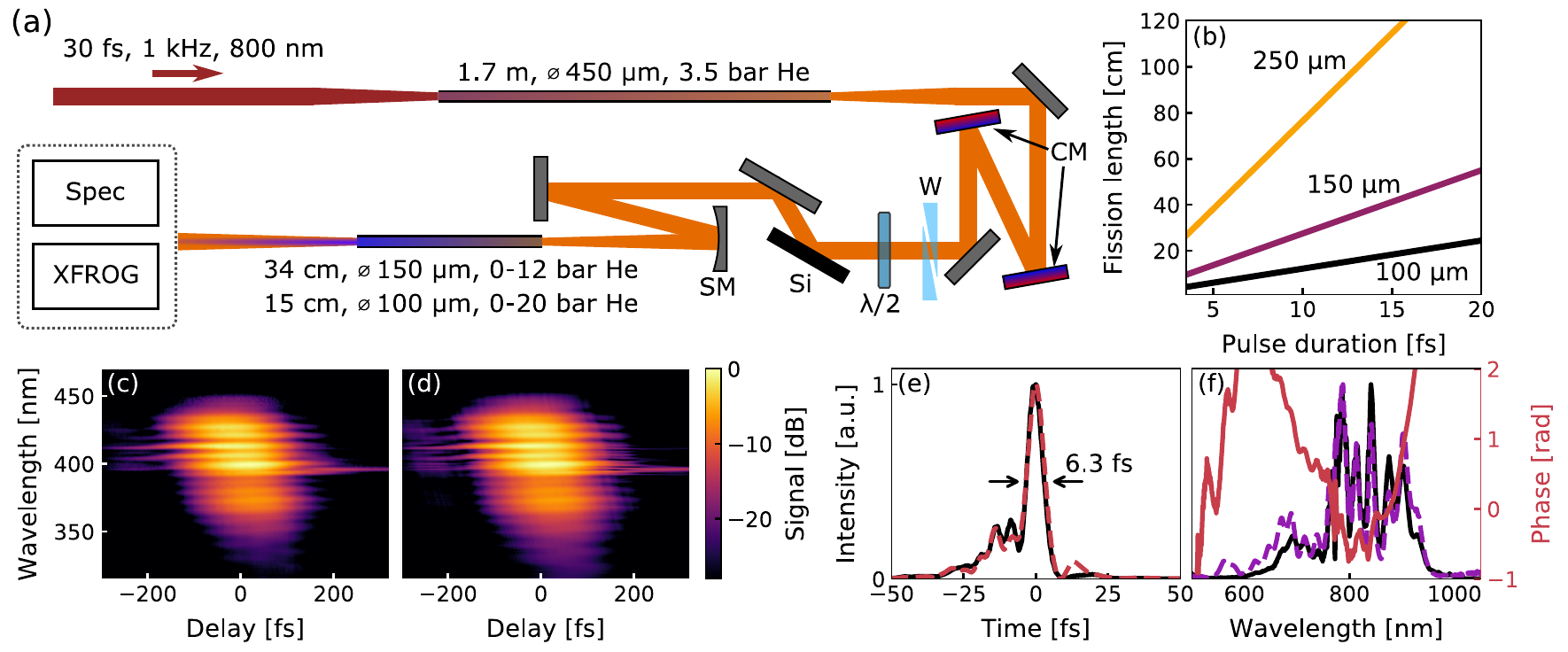}
    \caption{(a) Experimental layout: \SI{30}{\fs} pulses are spectrally broadened in helium-filled HCF and compressed by chirped mirrors (CM) and glass wedges (W). A half-wave plate ($\lambda/2$) and Brewster-angle silicon plate (Si) serve as a variable attenuator. The pulses are then coupled into the second HCF (\SI{100}{\micro\meter} or \SI{150}{\micro\meter} core diameter), also filled with helium, with a spherical mirror (SM). (b) Fission length as a function of initial pulse duration for a peak intensity of \SI{1.5e14}{\watt\per\square\cm} in helium-filled HCF with a zero-dispersion wavelength of \SI{560}{\nm} for 100, 150 and \SI{250}{\micro\meter} core diameter, requiring 13, 5.9 and \SI{2.1}{\bar} of pressure respectively.
    (c-f) Time-domain ptychography of the pump pulse. (c) SFG-XFROG trace measured after the evacuated \SI{150}{\micro\meter} HCF. (d) Retrieved trace after 100 iterations of the rPIE. (e) Temporal profile of the pump pulse as obtained by numerical back-propagation, measured after the \SI{150}{\micro\meter} (black) and \SI{100}{\micro\meter} HCF systems (red dashed). (f) Retrieved power spectrum (black) and phase (red) as well as the externally measured power spectrum (purple dashed) for comparison.}
    \label{fig:exp_Lfiss}
\end{figure*}
The HCF length required to generate a dispersive wave is primarily determined by the distance over which soliton self-compression occurs. This is well approximated by the fission length $L_\mathrm{f}$,
\begin{equation}
    L_\mathrm{f} = \frac{L_\mathrm{d}}{N} = \sqrt{L_\mathrm{d} L_\mathrm{nl}}\,,
\end{equation}
where $N$ is the soliton order and $L_\mathrm{d}$ and $L_\mathrm{nl}$ are the dispersion and nonlinear lengths, describing the length scales of group-velocity dispersion (GVD) and self-phase modulation (SPM), respectively \cite{dudley_supercontinuum_2006}. Broadly similar soliton dynamics and RDW emission can be obtained with different parameters, provided that the soliton order and the zero-dispersion wavelength $\lambda_\mathrm{zd}$ remain the same. In particular, the spectral location of RDW emission is chiefly determined by the pump wavelength $\lambda_0$ and $\lambda_\mathrm{zd}$. With both of these fixed, the fission length in HCF scales with the core radius $a$ and the pump pulse duration $\tau$ as
\begin{equation}
    L_\mathrm{f} \propto \frac{a^2\tau}{\sqrt{I_0}} \,,
    \label{eq:Lf}
\end{equation}
where $I_0$ is the peak intensity of the incident pump pulse \cite{travers_high-energy_2019}. This intensity is limited by the need to avoid self-focusing and excessive ionisation \cite{travers_high-energy_2019}. The most effective way of reducing the required HCF length is therefore the use of smaller core diameters. However, since the propagation loss of HCF increases dramatically for smaller cores, with the loss length $L_\textsc{l}$ scaling as $L_\textsc{l} \propto a^3$, this strategy reduces the overall throughput or even precludes soliton dynamics entirely \cite{travers_high-energy_2019}. Therefore, shorter pump pulses are also required. Fig.\ref{fig:exp_Lfiss}(b) shows this scaling for a zero-dispersion wavelength of \SI{560}{\nm} in three different core diameters---100 and \SI{150}{\micro\meter} as used here, and \SI{250}{\micro\meter} as used for the first demonstration of soliton dynamics in HCF \cite{travers_high-energy_2019}. Soliton dynamics can be obtained in very compact HCF systems when using short pulses and small cores.

The experimental layout is shown in Fig.\ref{fig:exp_Lfiss}(a). A commercial titanium-doped sapphire amplifier delivers \SI{30}{\fs} pulses at a repetition rate of \SI{1}{\kilo\hertz}. They are spectrally broadened in a \SI{1.7}{\m} long stretched HCF of \SI{450}{\micro\meter} core diameter filled with \SI{3.5}{\bar} of helium and compressed by 12 reflections from chirped mirrors (PC70, Ultrafast Innovations) and a fused silica wedge pair. An achromatic waveplate (B.Halle) and Brewster-angle silicon plate form a variable attenuator. The compressed pulses are coupled into a second HCF of either \SI{150}{\micro\meter} core diameter and \SI{34}{\cm} length or \SI{100}{\micro\meter} core diameter and \SI{15}{\cm} length for soliton self-compression. Output spectra are collected using a combination of an integrating sphere and CCD spectrometer. This system has been calibrated for absolute spectral response, allowing the extraction of UV pulse energies directly from the spectra (this has been verified by comparison to direct energy measurements). The spectra are corrected for the loss caused by reflection from the uncoated MgF$_2$ exit window of the HCF system.

The driving pulse is characterised using time-domain ptychography in a sum-frequency generation cross-correlation frequency-resolved optical gating (SFG-XFROG) apparatus, which is placed after either of the second HCFs \cite{witting_time-domain_2016}. In this way, the pulse measurement includes any spectral variation of the attenuation or waveguide coupling. The measured traces are corrected for any spectral selectivity in the apparatus by projecting their frequency marginal onto the expected marginal as calculated from the spectra of the unknown pulse and the narrowband gate pulse used in the XFROG. The pulse retrieval consists of 100 iterations of the regularised iterative ptychographic engine (rPIE) \cite{maiden_further_2017} using the measured trace and the gate pulse spectrum as the input. The pulse at the entrance of both HCFs, \SI{6.3}{\fs} in duration (full width at half maximum (FWHM) of the intensity), is shown in Fig.~\ref{fig:exp_Lfiss}(e-f). It is obtained by numerically back-propagating the retrieved pulse, taking into account the air path, the gas cell exit window and the capillary dispersion.

\begin{figure}[t]
    \includegraphics{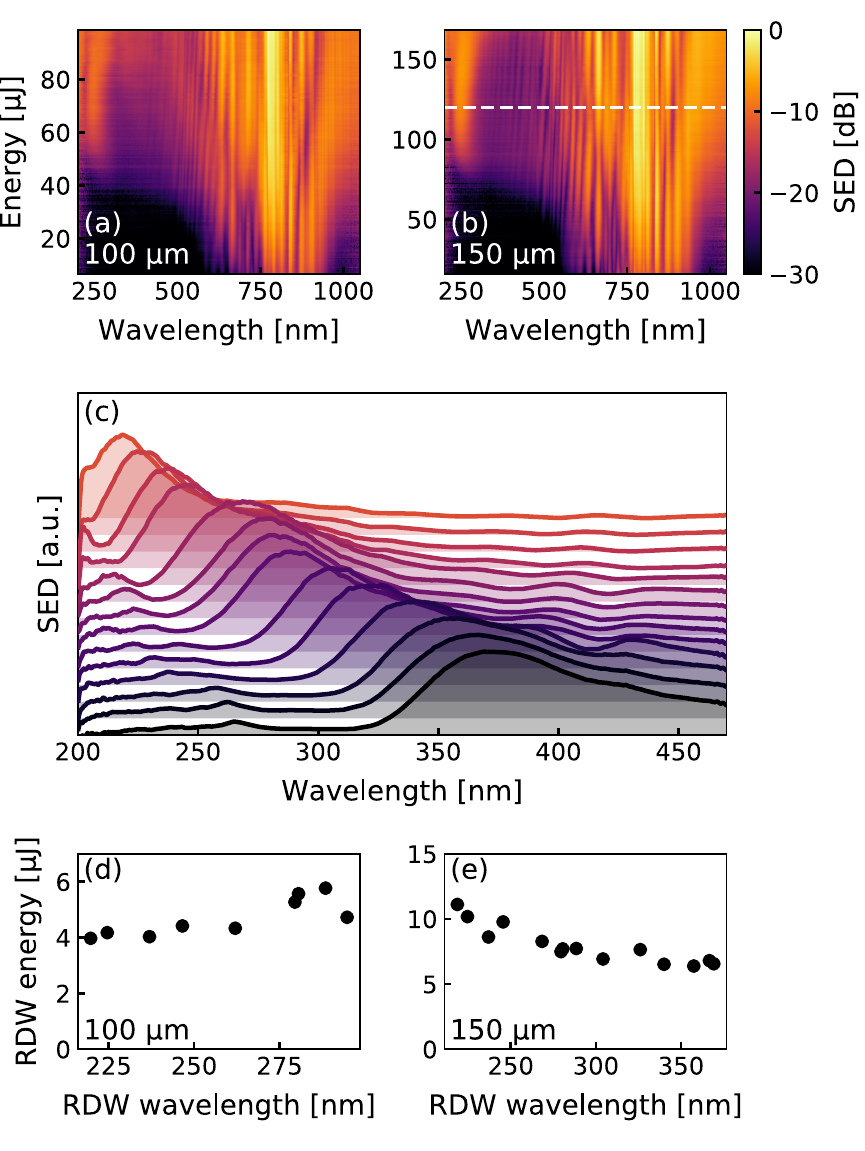}
    \caption{Dynamics of UV RDW emission in the small-core HCF. (a-b) Output spectra (corrected for Fresnel losses on the exit window) from (a) the \SI{100}{\micro\meter} HCF filled with \SI{12.3}{\bar} of helium and (b) the \SI{150}{\micro\meter} HCF filled with \SI{5.9}{\bar} of helium as the driving pulse energy (shown including coupling losses) is increased. SED: spectral energy density. (c) Tuneability of the RDW emission in the \SI{150}{\micro\meter} HCF. The lines show UV RDW spectra obtained with different pressures of helium (\SI{4.4}{\bar} at the shortest RDW wavelength to \SI{12.3}{\bar} at the longest), each normalised to its peak. (d-e) Energy in the UV spectrum as obtained from the (d) \SI{100}{\micro\meter} and (e) the \SI{150}{\micro\meter} (e) HCF.}
    \label{fig:escan}
\end{figure}

\begin{figure}[t]
    \includegraphics{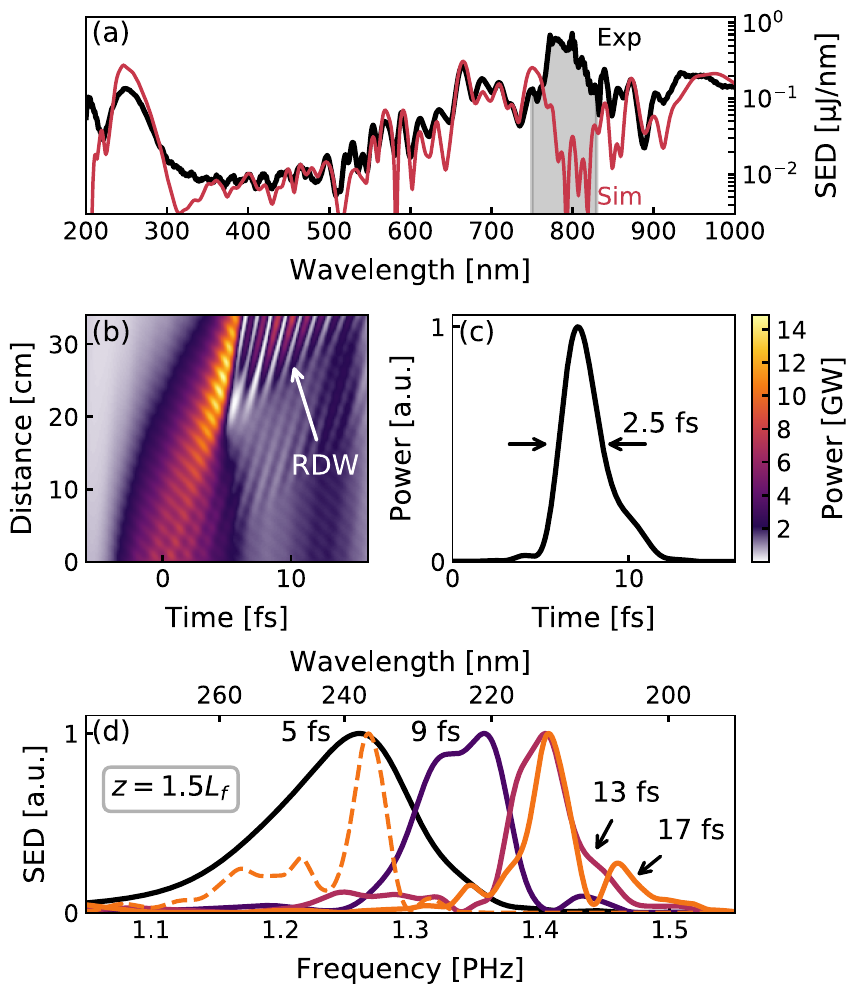}
    \caption{(a) Output spectrum from the \SI{150}{\micro\meter} HCF at \SI{120}{\micro\joule} of driving pulse energy, indicated by the white dashed line in Fig.~\ref{fig:escan}(b), both measured (black) and simulated (red), on an absolute energy scale. The grey shaded area indicates the spectral components likely due to poor pulse contrast, the energy of which is removed for the simulation. (b) Simulated propagation of the pulse. The RDW is emitted around \SI{25}{\cm} into the HCF. (c) Temporal profile of the RDW at \SI{30}{\cm} (the point of highest RDW peak power), obtained by band-pass filtering between 200 and \SI{300}{\nm}. (d) Solid lines: simulated RDW spectra for different driving pulse durations, keeping their peak power constant, extracted at 1.5 times their fission length in a lossless waveguide. The labels give the initial pulse duration of the driving pulse. Dashed line: RDW spectrum obtained with a \SI{17}{\fs} pulse with a reduced peak power.}
    \label{fig:sims}
\end{figure}

Fig.~\ref{fig:escan}(a-b) show the evolution of the output spectrum from the \SI{100}{\micro\meter} and \SI{150}{\micro\meter} HCF for one choice of $\lambda_\mathrm{zd}$, \SI{560}{\nm}, as the driving pulse energy is increased (the energy shown includes coupling losses). For these parameters, RDW emission is phase-matched at \SI{250}{\nm}. In the two different capillaries, the RDW appears at \SI{50}{\micro\joule} and \SI{85}{\micro\joule} of driving pulse energy respectively. That the dynamics are close to identical is a consequence of the general energy-scaling laws of soliton dynamics in HCF---by adjusting the gas pressure and energy such that $\lambda_\mathrm{zd}$ and $N$ are fixed, the same soliton dynamics can be obtained in HCF with different core diameter \cite{travers_high-energy_2019}. The scaling laws predict an energy ratio of $2.25$---that this is not reproduced exactly in our experiments (we use 1.7) is likely due to the higher propagation loss in the \SI{100}{\micro\meter} HCF, which is not accounted for in the scaling rules.

The phase-matching wavelength for RDW can be changed by adjusting the gas pressure, with higher pressure leading to RDW emission at longer wavelengths. Fig.~\ref{fig:escan}(c) shows this tuneability for the \SI{150}{\micro\meter} core diameter HCF, covering 218 to \SI{375}{\nm}. Near-identical spectra can be obtained at lower energy in the smaller core. Note that since HCF is free of the guidance resonances found in HC-PCF, the tuning is completely continuous without any gaps. The energy contained in the RDW is shown in Fig.~\ref{fig:escan}(d-e). In the \SI{150}{\micro\meter} HCF, 7 to \SI{11}{\micro\joule} are generated across the DUV. Even in the smaller HCF, the RDW pulse energy is several times larger than the highest reported in HC-PCF at all wavelengths we generate here \cite{kottig_generation_2017}. The conversion efficiency after correcting for coupling losses is between \SI{9}{\percent} and \SI{11}{\percent} at all wavelengths shown in Fig.~\ref{fig:escan}(c).

To gain more insight into the temporal structure and evolution of the dispersive wave, we have simulated the soliton self-compression using the single-mode unidirectional pulse propagation equation \cite{couairon_practitioners_2011, tani_multimode_2014}. The model includes the waveguide \cite{marcatili_hollow_1964} and gas \cite{borzsonyi_dispersion_2008} dispersion and the Kerr nonlinearity \cite{shelton_measurements_1994}. The effect of photoionisation \cite{geissler_light_1999} is included using the Perelomov-Popov-Terent'ev ionisation rate \cite{perelomov_ionization_1966}, however the ionisation fraction remains below $10^{-5}$ even at the highest peak intensity achieved in any of the simulations considered here (\SI{2.4e14}{\watt\per\square\cm}).

Fig.~\ref{fig:sims}(a) shows the comparison between the numerically modelled and experimentally measured output spectrum from the \SI{150}{\micro\meter} core diameter HCF at \SI{5.9}{\bar} pressure and \SI{120}{\micro\joule} driving pulse energy (indicated in Fig.~\ref{fig:escan}(b) by the white dashed line), using identical parameters to the experiment, including the measured driving pulse. The overall agreement is excellent, with the exception of a strong feature around the pump wavelength. We attribute this to poor pulse contrast of the initial \SI{30}{\fs} laser pulses, meaning that a significant part of the input energy does not contribute to the nonlinear interaction. The simulations take this into account by reducing the driving pulse energy from the measured value by \SI{30}{\micro\joule}, the energy contained in the shaded area in Fig.~\ref{fig:sims}(a). Note that this implies that even higher conversion efficiencies to the UV should be possible when using higher-quality driving pulses. On the other hand, that imperfect driving pulses give rise to high-quality RDW emission reflects the resilience of soliton dynamics in general, which can even arise from noise \cite{gouveia-neto_soliton_1989}. The feature around \SI{200}{\nm} in the experimental spectrum may be due to a dispersive wave in a higher-order mode \cite{tani_multimode_2014}, however it is likely dominated by the spectrometer noise being exaggerated by the spectral response calibration.

Given that we closely reproduce the spectrum of the RDW in the simulations, we can infer the general properties of the UV pulse generated in our experiment. Previous experiments in HC-PCF have shown that the RDW is emitted as a near transform-limited pulse with a duration as short as \SI{3}{\fs} when pumped with \SI{15}{\fs} pulses \cite{brahms_direct_2019}. From the simulations, we infer that the RDW pulses generated in these compact HCF systems are even shorter. As the propagation dynamics in Fig.~\ref{fig:sims}(b) show, the RDW is emitted around \SI{25}{\cm} into the HCF. At the point of highest peak power, \SI{30}{\cm} into the propagation, the RDW pulse (shown in Fig.~\ref{fig:sims}(c)) has a duration of \SI{2.5}{\fs}. Shortly before this point, at \SI{26}{\cm}, the pulse has reached \SI{50}{\percent} of its maximum energy but \SI{80}{\percent} of its maximum peak power with a pulse duration of only \SI{2}{\fs}. This is a conservative estimate of the RDW pulse duration, since the measured RDW spectrum is in fact even more broadband than the simulated one.

Differences in RDW spectra generated at the same wavelength but with different driving pulse durations have previously been investigated in numerical studies \cite{travers_ultrafast_2011}. Following the approach outlined in that work, we have simulated RDW emission for different driving pulse durations while keeping the normalised soliton order $S=N/\tau$ fixed (this is equivalent to constant peak power) by increasing the pulse energy for longer pulses. To allow a more direct comparison, the waveguide loss and thus the influence of the varying fission length is ignored in these simulations. 

Fig.~\ref{fig:sims}(d) shows the RDW spectra after propagating for 1.5 times the fission length for a \SI{150}{\micro\meter} core diameter HCF filled with \SI{5.9}{\bar} of helium when pumped with Gaussian-shaped pulses with $S=\SI{0.45}{\per\fs}$ (peak power of \SI{6.7}{\giga\watt}) and durations of 5, 9, 13 and \SI{17}{\fs}. Two main effects are visible: broader spectra for shorter pulses and shorter central wavelengths for longer pulses. It is well-known that the nonlinear contribution to the phase-matching causes a blue-shift of the RDW for higher initial peak power, since phase-matching to the RDW is primarily due to positive third-order dispersion \cite{joly_bright_2011, austin_dispersive_2006}. However, that this also occurs for pulses of the same initial peak power is more surprising and makes clear that the influence of the nonlinear phase is more complex.

Unless a pulse is so long in duration that modulational instability plays an important role, all initial pulses with the same peak power will compress to approximately the same minimum duration (the compressed pulse duration scales as $\tau_c\propto\tau/N\propto 1/S$), albeit with varying amounts energy contained in a low-power pedestal \cite{chen_nonlinear_2002}. Therefore the maximum peak power increases for longer, more energetic pulses---in our simulations, it is \SI{12}{\giga\watt} for the \SI{5}{\fs} pulse and \SI{20}{\giga\watt} for the \SI{17}{\fs} pulse. This increases the nonlinear phase contribution despite the identical initial peak power.

In addition, the quality of the self-compression (the relative amounts of energy in the self-compressed peak and the pedestal) depends on the soliton order $N$ rather than $S$ \cite{chen_nonlinear_2002}, so that longer pulses with the same peak power lead to a more complicated nonlinear phase landscape than shorter ones. The interplay between the peak power and the self-compression quality can be illustrated by reducing the energy of a longer pulse to shift the RDW to the same frequency as generated by a shorter one. The dashed line in Fig.~\ref{fig:sims}(d) shows the RDW obtained by reducing the initial peak power of the \SI{17}{\fs} pulse to \SI{2.3}{\giga\watt} ($S=\SI{0.27}{\per\fs}$, self-compressed peak power \SI{8.9}{\giga\watt}). At this power, the RDW is emitted at the same wavelength (\SI{238}{\nm}) as with the higher-power \SI{5}{\fs} driving pulse. The bandwidth is significantly narrower than generated by the \SI{5}{\fs} pulse, demonstrating that the initial pulse duration plays an important role in determining the shape of the RDW spectrum. Further detailed study is required to fully elucidate the relationship between pump pulse duration and RDW spectrum.

One of the strengths of RDW emission as a frequency conversion technique is that much shorter pulses than the driving pulse can be generated \cite{brahms_direct_2019}. The pulse duration scaling we have investigated here suggests that, like in other frequency up-conversion schemes, shorter driving pulses lead to shorter RDW pulses. Note, however, that this is due to a different mechanism than for e.g. third-harmonic generation (THG), where the up-converted pulse duration is directly determined by that of the driving pulse. In RDW emission, it is the evolution of the self-compressing driving pulse that causes the difference.

In conclusion, high-energy ultrashort pulses in the deep ultraviolet can be generated in simple hollow capillary fibres over distances as short as \SI{15}{\cm} and with driving pulses of less than \SI{100}{\micro\joule} energy. Pulse energies in the DUV of around \SI{10}{\micro\joule} and \SI{5}{\micro\joule} can be generated in \SI{150}{\micro\meter} and \SI{100}{\micro\meter} core diameter HCF, respectively, both of which are significantly higher than what has been achieved in HC-PCF. Unlike in HC-PCF, the achievable wavelength tuning range is not limited by guidance resonances. The RDWs generated with short driving pulses are more broadband, a fact which we attribute to the altered contribution of the nonlinear phase-matching. RDW emission in the VUV, as achieved in the first demonstration of HCF soliton dynamics \cite{travers_high-energy_2019}, will be straightforward to achieve. Due to the lower energy requirements, our approach will enable scaling of high-energy soliton dynamics in capillaries to much higher repetition rates than demonstrated so far, for instance using fibre lasers. The short driving pulses required are readily available either from existing pulse compression schemes or directly from lasers such as optical parametric chirped pulse amplification (OPCPA) systems. We expect that these capabilities in combination with the compactness we demonstrate here will find many applications in ultrafast science.

\section*{Funding}
This work was funded by the European Research Council (ERC) under the European Union's Horizon 2020 research and innovation program: Starting Grant agreement HISOL, No. 679649.

\bibliographystyle{unsrt}
\bibliography{bibliography}

\begin{thebibliography}{10}

\bibitem{joly_bright_2011}
N.~Y. Joly, J.~Nold, W.~Chang, P.~H{\"o}lzer, A.~Nazarkin, G.~K~L Wong,
  F.~Biancalana, and P.~St~J Russell.
\newblock Bright spatially coherent wavelength-tunable deep-{UV} laser source
  using an {Ar}-filled photonic crystal fiber.
\newblock {\em Physical Review Letters}, 106(20):203901, 2011.

\bibitem{kottig_generation_2017}
Felix K{\"o}ttig, Francesco Tani, Christian~Martens Biersach, John~C. Travers,
  and Philip~St.J. Russell.
\newblock Generation of microjoule pulses in the deep ultraviolet at megahertz
  repetition rates.
\newblock {\em Optica}, 4(10):1272, October 2017.

\bibitem{belli_vacuum-ultraviolet_2015}
Federico Belli, Amir Abdolvand, Wonkeun Chang, John~C Travers, and Philip St~J
  Russell.
\newblock Vacuum-ultraviolet to infrared supercontinuum in hydrogen-filled
  photonic crystal fiber.
\newblock {\em Optica}, 2(4):292--300, 2015.

\bibitem{ermolov_supercontinuum_2015}
A.~Ermolov, K.~F. Mak, M.~H. Frosz, J.~C. Travers, and P.~St~J Russell.
\newblock Supercontinuum generation in the vacuum ultraviolet through
  dispersive-wave and soliton-plasma interaction in a noble-gas-filled
  hollow-core photonic crystal fiber.
\newblock {\em Physical Review A}, 92(3), 2015.

\bibitem{travers_high-energy_2019}
John~C. Travers, Teodora~F. Grigorova, Christian Brahms, and Federico Belli.
\newblock High-energy pulse self-compression and ultraviolet generation through
  soliton dynamics in hollow capillary fibres.
\newblock {\em Nature Photonics}, April 2019.

\bibitem{dudley_supercontinuum_2006}
John~M. Dudley, Go{\"e}ry Genty, and St{\'e}phane Coen.
\newblock Supercontinuum generation in photonic crystal fiber.
\newblock {\em Reviews of Modern Physics}, 78(4):1135--1184, 2006.

\bibitem{witting_time-domain_2016}
T~Witting, D~Greening, D~Walke, P~Matia-Hernando, T~Barillot, J~P Marangos, and
  J~W~G Tisch.
\newblock Time-domain ptychography of over-octave-spanning laser pulses in the
  single-cycle regime.
\newblock {\em Opt. Lett.}, 41(18):4218--4221, 2016.

\bibitem{maiden_further_2017}
Andrew Maiden, Daniel Johnson, and Peng Li.
\newblock Further improvements to the ptychographical iterative engine.
\newblock {\em Optica}, 4(7):736, July 2017.

\bibitem{couairon_practitioners_2011}
A.~Couairon, E.~Brambilla, T.~Corti, D.~Majus,
  O.~de~J.~Ram{\'i}rez-G{\'o}ngora, and M.~Kolesik.
\newblock Practitioner{\textquoteright}s guide to laser pulse propagation
  models and simulation.
\newblock {\em The European Physical Journal Special Topics}, 199(1):5--76,
  2011.

\bibitem{tani_multimode_2014}
Francesco Tani, John~C Travers, and Philip St.J.~Russell.
\newblock Multimode ultrafast nonlinear optics in optical waveguides: numerical
  modeling and experiments in kagom{\'e} photonic-crystal fiber.
\newblock {\em Journal of the Optical Society of America B}, 31(2):311, 2014.

\bibitem{marcatili_hollow_1964}
E.~A.~J. Marcatili and R.~A. Schmeltzer.
\newblock Hollow {Metallic} and {Dielectric} {Waveguides} for {Long} {Distance}
  {Optical} {Transmission} and {Lasers}.
\newblock {\em Bell System Technical Journal}, 43(4):1783--1809, July 1964.

\bibitem{borzsonyi_dispersion_2008}
A.~B{\"o}rzs{\"o}nyi, Z.~Heiner, M.~P. Kalashnikov, A.~P. Kov{\'a}cs, and
  K.~Osvay.
\newblock Dispersion measurement of inert gases and gas mixtures at 800 nm.
\newblock {\em Applied Optics}, 47(27):4856, September 2008.

\bibitem{shelton_measurements_1994}
David~P. Shelton and Julia~E. Rice.
\newblock Measurements and {Calculations} of the {Hyperpolarizabilities} of
  {Atoms} and {Small} {Molecules} in the {Gas} {Phase}.
\newblock {\em Chemical Reviews}, 94(1):3--29, 1994.

\bibitem{geissler_light_1999}
M.~Geissler, G.~Tempea, A.~Scrinzi, M.~Schn{\"u}rer, F.~Krausz, and T.~Brabec.
\newblock Light {Propagation} in {Field}-{Ionizing} {Media}: {Extreme}
  {Nonlinear} {Optics}.
\newblock {\em Physical Review Letters}, 83(15):2930--2933, October 1999.

\bibitem{perelomov_ionization_1966}
A~M Perelomov, V~S Popov, and M~V Terent~'ev.
\newblock Ionization of atoms in an alternating electric field.
\newblock {\em Soviet Physics JETP}, 23(50):1393--1409, 1966.

\bibitem{gouveia-neto_soliton_1989}
A.~S. Gouveia-Neto and J.~R. Taylor.
\newblock Soliton evolution from noise bursts (optical fibres).
\newblock {\em Electronics Letters}, 25(11):736--737, May 1989.

\bibitem{brahms_direct_2019}
Christian Brahms, Dane~R. Austin, Francesco Tani, Allan~S. Johnson, Douglas
  Garratt, John~C. Travers, John W.~G. Tisch, Philip~St.J. Russell, and Jon~P.
  Marangos.
\newblock Direct characterization of tuneable few-femtosecond dispersive-wave
  pulses in the deep {UV}.
\newblock {\em Optics Letters}, 44(4):731, February 2019.

\bibitem{travers_ultrafast_2011}
John~C Travers, Wonkeun Chang, Johannes Nold, Nicolas~Y Joly, and Philip St.
  J.~Russell.
\newblock Ultrafast nonlinear optics in gas-filled hollow-core photonic crystal
  fibers [{Invited}].
\newblock {\em Journal of the Optical Society of America B}, 28(12):A11--A26,
  2011.

\bibitem{austin_dispersive_2006}
Dane~R Austin, C~Martijn de~Sterke, Benjamin~J Eggleton, and Thomas~G Brown.
\newblock Dispersive wave blue-shift in supercontinuum generation.
\newblock {\em Optics express}, 14(25):11997--12007, 2006.

\bibitem{chen_nonlinear_2002}
Chia-Ming Chen and Paul~L. Kelley.
\newblock Nonlinear pulse compression in optical fibers: scaling laws and
  numerical analysis.
\newblock {\em Journal of the Optical Society of America B}, 19(9):1961,
  September 2002.

\end{thebibliography}

\end{document}